\documentclass[10pt, conference]{IEEEtran}

\usepackage{tikz}
\usepackage{amsmath}
\usepackage{tcolorbox}
\usepackage{booktabs}
\usepackage{array}
\usepackage{listings}
\usepackage{xcolor}
\usepackage{footnote}
\usepackage{pdfpages}
\usepackage{float}
\usepackage{afterpage}
\usepackage{url}
\usepackage{multirow}
\usepackage{makecell}
\usepackage{filecontents}
\usepackage{hyperref}
\usetikzlibrary{positioning}
\usepackage{graphicx}
\usepackage{cite}

\definecolor{KWColor}{rgb}{0.37,0.08,0.25}
\definecolor{CommentColor}{rgb}{0.133,0.545,0.133}
\definecolor{StringColor}{rgb}{0,0.126,0.941}
\newcommand{\filledcircle}{\tikz\fill[black] (0,0) circle (0.1cm);}
\newcommand{\emptycircle}{\tikz\draw (0,0) circle (0.1cm);}

\newcommand{\halfcircle}{%
    \tikz{
    \draw[fill=black] (0,0) -- (0,0.1cm) arc[start angle=90, end angle=270, radius=0.1cm] -- cycle;
    \draw[fill=white] (0,0) -- (0,-0.1cm) arc[start angle=270, end angle=450, radius=0.1cm] -- cycle;
    \draw (0,0) circle (0.1cm);
    }
}
\usepackage{filecontents}
\usepackage{url}
\usepackage{multirow}
\usepackage{makecell}

\definecolor{Cerulean}{RGB}{0,123,167}  
\definecolor{lightblue}{RGB}{201, 236, 250}  
\definecolor{mediumblue}{RGB}{121, 211, 243}  
\definecolor{darkblue}{RGB}{0, 173, 231}

\colorlet{tilecolor}{Cerulean}

\begin{document}
\pagestyle{plain}

\date{}

\title{A First Look at Privacy Risks of Android Task-executable Voice Assistant Applications}

\author{
\IEEEauthorblockN {    
   Shidong Pan$^{a,b}$,       
   Yikai Ge$^{c}$,
   Xiaoyu Sun$^{c,\dagger}$
}
\vspace{5pt}
\IEEEauthorblockA {
  $^{a}$\textit{New York University}
  $^{b}$\textit{Columbia University}
  $^{c}$\textit{Washington University}
  $^{d}$\textit{Australian National University}
}
  $^\dagger$\textit{Corresponding author} (xiaoyu.sun1@anu.edu.au)
}

\maketitle

\begin{abstract}
With the development of foundation AI technologies, task-executable voice assistants (VAs) have become more popular, enhancing user convenience and expanding device functionality.
Android task-executable VAs are applications that are capable of understanding complex tasks and performing corresponding operations.
Given their prevalence and great autonomy, there is no existing work examine the privacy risks within the voice assistants from the task-execution pattern in a holistic manner. 
To fill this research gap, this paper presents a user-centric comprehensive empirical study on privacy risks in Android task-executable VA applications.
We collect ten mainstream VAs as our research target and analyze their operational characteristics. 
We then cross-check their privacy declarations across six sources, including privacy labels, policies, and manifest files, and our findings reveal widespread inconsistencies.
Moreover, we uncover three significant privacy threat models: (1) privacy misdisclosure in mega apps, where integrated mini apps such as Alexa skills are inadequately represented; (2) privilege escalation via inter-application interactions, which exploit Android's communication mechanisms to bypass user consent; and (3) abuse of Google system applications, enabling apps to evade the declaration of dangerous permissions.
Our study contributes actionable recommendations for practitioners and underscores broader relevance of these privacy risks to emerging autonomous AI agents.
\end{abstract}

\section{Introduction}

Voice assistants (VA) constitute an indispensable technological interface for individuals with visual impairments or those unable to manipulate traditional application interfaces, serving as a critical conduit for accessing digital services and information.
For instance, Apple's Siri\footnote{\url{https://www.apple.com/au/siri/}}, Amazon's Alexa\footnote{\url{https://www.alexa.com/}}, and Google Assistant\footnote{\url{https://assistant.google.com/}} have revolutionized how users interact with their devices, allowing for hands-free operation of smartphones, smart home devices, and even vehicles~\cite{murad2019revolution,clark2019state}.
These intelligent systems represent a significant advancement in human-computer interaction, offering substantial benefits for users with disabilities. 
The integration of them into everyday devices has the potential to significantly ameliorate the digital divide experienced by individuals with diverse accessibility needs, fostering greater independence and societal inclusion.

The rapid advancement of artificial intelligence and associated technologies has propelled VAs to the forefront of human-computer interaction paradigms. 
Voice-based assistants like Alexa, Siri, and Google Assistant have evolved beyond simple chatbot functionalities to become sophisticated, task-oriented platforms integral to daily life. 
These advanced VAs leverage cutting-edge technologies such as Computer Vision and Natural Language Processing to comprehend and execute complex commands across various domains~\cite{bolton2021security, acosta2022survey, liao2020measuring, wu2025assistants, lin2025mind}. 
\textbf{Task-executable VAs} represent a significant leap forward in functionality. Unlike their predecessors, which were primarily limited to answering queries and performing basic tasks, these modern VAs can interact with multiple applications, control smart home devices, make purchases, and even perform complex multi-step operations. 
This enhanced capability allows them to seamlessly integrate into users' digital ecosystems, offering a more intuitive and efficient way to interact with technology. 
However, the more system access granted and increased autonomy to task-executable VAs raise critical questions regarding data privacy.

The unpredictability of these devices has been further illustrated by several notable incidents. 
A family in Portland, Oregon, discovered that their Amazon Echo device had not only recorded a private conversation without their knowledge but also sent the recording to a random contact in their address book. 
Amazon later explained that the device had misinterpreted background conversation as a series of commands, leading to the accidental recording and sharing of the conversation\footnote{\href{https://www.theguardian.com/technology/2018/may/24/amazon-alexa-recorded-conversation}{theguardian.com/technology/amazon-alexa-recorded-conversation}}. 
These privacy concerns raised attentions from legislation, particularly with regard to the enforcement of privacy laws and regulations.
A salient example illustrating these concerns is Amazon’s \$25 million settlement with the U.S. Federal Trade Commission for violations of the Children’s Online Privacy Protection Act (COPPA)\footnote{\href{https://www.nytimes.com/2023/05/31/technology/amazon-25-million-childrens-privacy.html}{nytimes.com/technology/amazon-25-million-childrens-privacy}}. 
The case revealed that Amazon retained children’s voice recordings and location data despite deletion requests, exposing the disconnect between declared privacy policies and actual practices~\cite{bbcamazon2023}.
Existing studies mainly focus on traditional VAs~\cite{natatsuka2019poster} or on certain ecosystems~\cite{xie2022scrutinizing}, thus, a holistic study about Android task-executable VAs is pressingly needed.

In this paper, we echo the practical need to conduct a holistic  privacy risk examination and analysis of Android task-executable VA applications.
We begin by presenting a clear definition of the task-executable VAs and examining the current status quo of VAs in the market.
We collect ten mainstream VAs as our research target by certain inclusion and exclusion criteria.
Subsequently, we perform a rigorous characterization of these VAs along eight evaluative dimensions, encompassing both operational efficiency and reliability. During this analysis, we identify substantial inconsistencies between the privacy disclosures presented to users and the actual behaviors observed through empirical testing. Motivated by this discovery, we conduct a holistic cross-source privacy examination across six information sources: privacy labels from the market, privacy policies from the market, results from Android permission checkers, declaration from APK manifest files, Android system settings, and the actual usage.
Our findings reveal that privacy labels frequently fail to disclose permissions that are otherwise detectable through static analysis or explicitly declared within the manifest files. Among the applications analyzed, Google Assistant and Amazon Alexa exhibit the most significant inconsistencies across data sources. In contrast, Voice Search (V.K.D) and Voice Search Assistant demonstrate relatively better alignment between declared and observed behaviors.

Our in-depth analysis of the internal workflows and interaction mechanisms of these VAs uncovers three prominent privacy threat models: (1) Privacy Misdisclosure in Mega Applications (Section~\ref{question2}); (2) Privilege Escalation via Inter-Application Interactions (Section~\ref{question3}); and (3) Abuse of Google System Applications (Section~\ref{question4}).
Specifically, we first pinpoint that the privacy notices of mega apps often overlook the disclosure of privacy usage related to their mini apps (e.g., Alexa skills) or integrated functionalities. 
Then, we reveal a sophisticated privilege escalation attack model that exploits inter-app interaction mechanisms in task-executable VAs. This attack model leverages the inherent pathways between applications to escalate privileges without user consent.
Moreover, we discover a sophisticated privilege escalation pattern that VAs might exploit the Google system applications to avoid disclose dangerous permissions in their own privacy notices.
Lastly, we summarize recommendations to practitioners and discuss the implications to the autonomous AI Agents.

\textbf{Contribution.}
In summary, this study presents a comprehensive analysis of privacy risks in Android task-executable voice assistants, uncovering privacy disclosure issues and critical privilege escalation thread models. Our findings provide actionable insights for practitioners and could be generalized to broader autonomous AI Agents.


\section{Status Quo of Task-executable VAs}\label{section1}

%

\begin{table}[t]
\centering
  \caption{Examples of non-task-executable VA applications that are excluded based on selection criteria.}
  \label{table:excluded_chatbots}
  \resizebox{0.99\linewidth}{!}{%
  \begin{tabular}{c|llr} 
    \toprule
    \textbf{\makecell[c]{Non \\ Task-executable}} &  \textbf{App Name} & \textbf{Developer} & \textbf{Downloads} \\
    \midrule

\multirow{5}{*}{Chatbot} 
    &ChatGPT &  OpenAI & 50M+\\
    & Pi  & Inflection AI & 100K+\\
    &ChatOn & AIBY & 5M+\\
    & Luzia & Luzia & 1M+ \\
    & Chatbot AI & Codespace Dijital & 5M+ \\
    \midrule
    \multirow{4}{*}{\makecell[c]{Text-Voice \\ Translate Assistant}}
    &AI Call Assistant \& Screener & Call Assistant Inc & 50K+  \\
 &Speech Assistant AAC  & ASoft.nl & 500K+\\
 &Lookout-Assisted Vision  & Google & 500K+\\
 &Speechify & Speechify& 1M+\\
 \midrule
    \multirow{2}{*}{\makecell[c]{Smart Home \\ Assistant}} &Home Assistant & Home Assistant & 1M+  \\
 & Vision - Smart Voice Assistant & Turbo Trade S.A.& 100K+  \\
 \midrule
  Spam Call Blocker & Truecaller: Spam Call Blocker  & Truecaller & 1B+  \\

  \bottomrule
\end{tabular}
}%
\vspace{-5pt}
\end{table}

%
%
\begin{table*}[t]
\centering
  \caption{Collected task-executable VA applications from the Google Play Store.}
  \label{table:chatbot_search}
  \resizebox{0.98\linewidth}{!}{%
  \begin{tabular}{c|r|lllr} 
    \toprule
    \textbf{Type} &  \textbf{\#} &  \textbf{App Name} & \textbf{Package Name} & \textbf{Developer} & \textbf{Downloads} \\
    \midrule

\multirow{10}{*}{\makecell[c]{\textbf{Task-executable} \\ Voice Assistants}} & 1 & Google Assistant & com.google.android.apps.googleassistant & Google & 1B+\\ 
 &2 & Voice Access & com.google.android.apps.accessibility.voiceaccess & Google & 100M+  \\
 &3 & Amazon Alexa & com.amazon.dee.app & Amazon Mobile  & 100M+  \\
 &4 & Ultimate Alexa Voice Assistant & com.customsolutions.android.alexa & Custom Solutions & 5M+  \\
 &5 & oice Search (UXAPPS) & ru.yvs & UXAPPS LTD & 5M+ \\
 &6 & Voice Search Assistant & com.combo.voiceassistant & Standard Applications & 1M+ \\
 &7 & Voice Search: Search Assistant & com.prometheusinteractive.voice\_launcher &Prometheus Interactive & 10M+ \\
 &8 & Voice Search (Preeti Devi) & com.appybuilder.onlinehelp3011.BestVoiceSearch & Preeti Devi & 1M+\\
 &9 & Voice Search (V.K.D) & com.onlinehelp3011.VoiceSearch &V.K.D & 100k+\\
 &10 & Voice Search (AE) & jp.gr.java\_conf.mamama.voicesearchcw & AE App World & 100K+\\
  \bottomrule
\end{tabular}
}%
\end{table*}
%
%

Breakthroughs in AI significantly contribute to the advancement of VAs. 
From the initial simple instruction executors to the current provider of intelligent dialogue and personalized services, VAs have undergone significant development~\cite{terzopoulos2020voice, hoy2018alexa, kepuska2018next}.
However, this increasing sophistication and expanded functional scope also give rise to critical privacy concerns that demand thorough analysis. 
Prior to diving into to-be-discussed privacy risks, we believe it is essential to establish a comprehensive understanding of task-executable VA applications, particularly with respect to their core characteristics and operational capabilities.

\subsection{Task-executable VA Applications}

Task-executable voice assistants are commonly AI-empowered applications that are capable of understanding complex tasks and performing corresponding operations on mobile phones or connected IoT devices.
For example, Amazon Alexa can invoke the TED Talks app on the phone by the voice command ``Alexa, ask TED Talks to play a talk about technology''.
Normally, task-executable VAs take voice input and translate it to text through Automatic Speech Recognition techniques. 
This text is then processed with Natural Language Understanding (NLU) which is normally powered by pre-trained Large Language Models (LLMs), producing a series of operations to complete the specific task~\cite{hoy2018alexa, cheng2022personal}.
After, they ``automatically'' command the phone or IoT devices to perform operations (e.g., open the TED Talks app) and then convert the textual responses back to speech using Text-to-Speech method.
Task-executable VAs extend far beyond the mere provision of query-based responses, but they are capable of initiating and executing real-world operations in response to user voice commands. 
This operational capacity differentiates them substantially from conventional voice-based applications with limited interactivity.

To delineate the scope of task-executable VAs, it is essential to identify those that do not fall within this category. As summarized in Table~\ref{table:excluded_chatbots}, non-task-executable voice assistant applications can be empirically classified into four primary categories: ``Chatbot'', ``Text-Voice Translate Assistant'', ``Smart Home Assistant'', and ``Spam Call Blocker''. 
Among these categories, a significant portion of applications primarily function as chatbots. 
These applications are predominantly characterized by their ability to provide extended text-based responses and output by reading this content aloud.
They cannot automate operations or perform complex tasks such as searching online by voice inputs.
Other applications that are non task-executable can be categorized as different assistant tools, and each designed to deliver specific functionalities. 
Notably, some of these tools provide visual assistance by converting voice to text and text to voice, thereby enhancing accessibility for users. Additionally, certain applications specialize in blocking spam calls, thereby protecting users from unwanted communications.

\subsection{Task-executable VA Application Collection}
We first collect those voice assistants in the Google Play app store which is the most largest and the most accessible app market.
The Google Play app store is often more transparent and friendly to academic research~\cite{pan2023toward, pan2024hope, pan2024trap}.
To ensure a comprehensive and controlled experimental environment, we utilize the searching feature of the Google Play app store to collect target applications.
We use an initial set of search terms, including ``$AI \, Assistant$'', ``$Voice \, Assistant$'',``$Smart\, Assistant$'', ``$AI\, Agent$'', \\``$Smart\, Agent$'',``$Voice\, Agent$'', \textit{``Personal
Assistant''}.
To evaluate search results, we manually select the top 20 results for each search query based on their default ranking.
In addition to the direct results returned by the Google Play, we also collect \textit{similar applications} that automatically recommended by the Google Play.
After obtaining the initial set of results and removing duplicates, we narrow down the task-executable voice assistant applications based on the following four criteria: 1)The application must support activation via voice commands as inputs; 2) The application must provide immediate verbal feedback upon receiving a command; 3) The application must execute meaningful operations beyond simple reading textual responses; 4) The application should have at least 50,000 downloads (\textit{i.e.}, installs) on Google Play. 
This method allows us to maintain a controlled and relevant set of applications for our experiments.
Table~\ref{table:chatbot_search} presents 10 identified task-executable voice assistant applications as the research targets in this paper.

\begin{table*}
  \centering
  \caption{The characterization of 10 voice assistants on seven different dimensions.}
  \label{table:characterization}
  \resizebox{0.98\linewidth}{!}{%
  \begin{tabular}{r|cc|ccc|ccc} 
    \toprule
    \textbf{Name} & \rotatebox{45}{\textbf{Freeware}} & \rotatebox{45}{\textbf{Registration}} & \rotatebox{45}{\textbf{\shortstack{Battery\\Consumption}}} & \rotatebox{45}{\textbf{\shortstack{Data\\Consumption}}} & \rotatebox{45}{\textbf{\shortstack{RAM\\Occupation}}} & \rotatebox{45}{\textbf{\shortstack{Response\\Time}}}& \rotatebox{45}{\textbf{\shortstack{Recognition\\Accuracy}}} & \rotatebox{45}{\textbf{Robustness}} \\
    \midrule
    Google Assistant & Free & Not Required & \filledcircle & \filledcircle & \halfcircle & \filledcircle & \halfcircle & \halfcircle \\ 
    Voice Access & Free & Not Required & \filledcircle & N/A & \filledcircle & \emptycircle & \filledcircle & \filledcircle \\
    Amazon Alexa & Free & Required & \halfcircle & \halfcircle & \emptycircle & \filledcircle & \filledcircle & \filledcircle \\
    Ultimate Alexa Voice Assistant & Free\&Paid & Required & \halfcircle & \halfcircle & \emptycircle & \emptycircle & \filledcircle & \filledcircle \\
    Voice Search (UXAPPS) & Free & Not Required & \halfcircle & \filledcircle & \halfcircle & \emptycircle & \filledcircle & \filledcircle \\
    Voice Search Assistant & Free & Not Required & \halfcircle & \emptycircle & \emptycircle & \halfcircle & \filledcircle & \filledcircle \\
    Voice Search: Search Assistant & Free\&Paid & Not Required & \halfcircle & \emptycircle & \halfcircle & \emptycircle & \filledcircle & \filledcircle \\
    Voice Search (Preeti Devi) & Free & Not Required & \emptycircle & \emptycircle & \emptycircle & \emptycircle & \filledcircle & \filledcircle \\
    Voice Search (V.K.D) & Free & Not Required & \emptycircle & \emptycircle & \halfcircle & \emptycircle & \filledcircle & \filledcircle \\
    Voice Search (AE) & Free & Not Required & \halfcircle & \halfcircle & \halfcircle & \emptycircle & \filledcircle & \filledcircle \\
  \bottomrule
\end{tabular}
}%
\end{table*}

%
%
\subsection{Characterization}
Many studies have intensively investigated the voice assistants~\cite{xie2022scrutinizing,lau2018alexa,huang2020amazon,alrawi2019sok,liao2024understanding}, whilst few have investigated the unique challenges brought by the task-executable feature.
We first conduct an empirical investigation to assess their capabilities and understand their characteristics, facilitating our follow-up research toward their privacy risks.
The comprehensive assessment is conducted on identified voice assistants along eight dimensions, and results are shown in the Table~\ref{table:characterization}.
The first two dimensions are their basic information; the next three dimensions target the efficiency of usage; and the last three dimensions focus on the reliability of voice assistants. 
To ensure the consistency of evaluation, all evaluation is conducted on a Google’s Pixel 7a, with 8GB memory and Android 14 operating system.
We specifically define the eight dimensions as follows:

\textbf{(1) Freeware}.
This aspect denotes whether the VA application is available as a free service, a paid service, or a combination of both. 
We categorize the software into one of three groups: completely free, requiring payment for use, or offering both free and paid tiers. 
This evaluation helps determine the accessibility of the assistant for users with different budget considerations.

\textbf{(2) Registration}.
In this criterion, we evaluate whether the VA application requires user registration before use. 
We categorize the software as either requiring mandatory registration or allowing access without any registration process. This assessment is important for understanding the setup complexity and privacy implications for users.

\textbf{(3) Battery Consumption}.
Users of VA applications frequently interact with their mobile phones, making battery consumption a primary concern for long-term usage.
To simulate real-world usage, we use ten testing input commands and interact with the target VA once per minute for a total of ten minutes. 
Battery consumption is measured from the opening of the specific VA application to the end of the last response. 
We observe they may utilize third-party applications (e.g., Google Search, speech recognition) to provide functions. 
The battery consumption of these third-party applications is included in the total battery consumption of the voice assistant, as they are integral to the interaction. 
Consumption was monitored and collected by using AccuBattery\footnote{\url{https://accubatteryapp.com/}}, and the unit of measurement is milliampere-hours (mAh). 
The VA applications are scored as (\filledcircle) if their battery consumption is less than 20 mAh, (\halfcircle) if between 20 to 40 ; and (\emptycircle) if greater than 40.

\textbf{(4) Data Consumption}. 
Some VAs provide services using the Internet to access database and online AI models, for purposes such as task planning, content searching, or accurate voice recognition. 
To monitor their mobile data usage during the ten testing input commands (as used for battery consumption), we employ GlassWire\footnote{\url{https://www.glasswire.com/}}. 
The data usage of third-party applications triggered by the voice assistant is also included in the total data consumption.
The data is calculated as the average usage across the ten tested commands, with the unit of measurement being kilobytes (KB). 
The VA applications are scored as (\filledcircle) if they only consumed data less than 500 KB, (\halfcircle) for 500 KB to 1,000 KB, and (\emptycircle) for more than 1,000 KB.
    
\textbf{(5) RAM Occupation}. 
The RAM occupation of applications on mobile phones plays an important role in stability.
During the battery consumption test, we monitor the running services using the Android Developer tool and record the peak RAM occupation. 
The RAM usage of third-party applications triggered by the voice assistant is also included.
The measurement unit for RAM occupation is megabytes (MB). 
We use (\filledcircle) indicates below 200 MB and (\halfcircle) between 200MB to 400 MB; (\emptycircle) for 400 MB above.

\textbf{(6) Response Time}. 
Response time of voice assistants is a critical factor of user experience.
The processing time can vary depending on the complexity of the tasks and capability of driven models. 
Here, we compute the average response time of ten test cases and the measurement unit is second(s). 
The response time between 0 to 0.5 second is marked as (\filledcircle), (\halfcircle) for 0.5 to 1 second, and (\emptycircle) for more than 1 second.

\textbf{(7) Recognition Accuracy}.
The speech recognition accuracy is essential to effectively execute the voice commands. 
When the VA is activated and the users begin speaking, the recognized text are commonly displayed on the screen. 
To evaluate the accuracy, we conduct the assessment with the same ten commands. 
For each command, we record whether the speech recognition accurately transcribed the spoken words into text.
The scoring is based on the number of commands correctly recognized out of ten. 
(\filledcircle) for all ten correct transcriptions, (\halfcircle) for five to ten correct transcriptions, and (\emptycircle) for less than five. 

\begin{table*}[t]
\centering
\caption{The mapping between Google Data Safety types, Android permissions, and testing commands.}
\label{tab:data-safety-permissions}
\resizebox{0.9\linewidth}{!}{
\begin{tabular}{l|l|l|l}
\toprule
\textbf{\shortstack{Privacy Labels \\ (1) }} & \textbf{\shortstack{Relevant Android Permissions \\ (2, 3, 4) }} & \textbf{\shortstack{Settings \\ (5) }} & \textbf{\shortstack{Testing Commands \\ (6) }}\\
\midrule
Approximate location & ACCESS\_COARSE\_LOCATION & Location & -\\
Precise location & ACCESS\_FINE\_LOCATION & Location & Show my location\\
\midrule
Phone number & READ\_PHONE\_STATE & PersistentID & - \\
\midrule
Fitness info & ACTIVITY\_RECOGNITION & Sensor & -\\
\midrule

SMS or MMS & READ\_SMS & SMS & Send message to ``myself", what is going on. \\
\midrule

Photos & READ\_EXTERNAL\_STORAGE & Storage & - \\
 &CAMERA & Camera & -\\
Videos & READ\_EXTERNAL\_STORAGE & Storage & - \\
 & CAMERA  & Camera & -\\
 \midrule
Voice or sound recordings & RECORD\_AUDIO & Microphone & - \\
 & READ\_VOICEMAIL & Microphone & Read voice mail.\\
Music files & READ\_EXTERNAL\_STORAGE & Storage & - \\
Other audio files & READ\_EXTERNAL\_STORAGE & Storage & - \\
\midrule
Files and docs & READ\_EXTERNAL\_STORAGE  & Storage & - \\
 & MANAGE\_EXTERNAL\_STORAGE & Storage & Manage external storage.\\
\midrule
Calendar events & READ\_CALENDAR & Calendar & - \\
 & WRITE\_CALENDAR & Calendar & Add meeting at 2 o'clock on tomorrow.\\
\midrule
Contacts & READ\_CONTACTS & Contacts & Call ``myself" in contacts. \\
 & WRITE\_CONTACTS & Contacts & Add contacts with phone number 222 an name is Dam. \\
  & GET\_ACCOUNTS & Contacts & Show accounts number. \\

\midrule
Installed apps & QUERY\_ALL\_PACKAGES & - & - \\
\midrule

Diagnostics & BATTERY\_STATS  & - & Check battery stats. \\
\midrule
Device or other IDs & READ\_PHONE\_STAT & -  & - \\
 & BLUETOOTH  & - & Pair bluetooth with headphone. \\
\bottomrule
\end{tabular}
}
\end{table*}

\textbf{(8) Robustness}.
Speech recognition accuracy across diverse accents is a critical metric for evaluating the robustness of VAs. 
To assess this capability, we employ AI-generated voice software, 
LOVO\footnote{\url{https://lovo.ai/}}, 
to simulate three English accents: Chinese English, Singaporean English, and Indian English.  
The ten testing commands are vocalized using the AI voice generators and subsequently played to the VAs. 
The accuracy of speech recognition was determined by analyzing the text results displayed on the screen, and we use the same metrics as the recognition accuracy.



As shown in Table~\ref{table:characterization}, the comparison between VA applications across eight different dimensions reveals varying levels of performance.
Regarding freeware status, most assistants are free, with a few like Ultimate Alexa Voice Assistant offering both free and paid versions. As for registration requirements, only Amazon Alexa and Ultimate Alexa Voice Assistant necessitate registration, while the others do not require it.
Google Assistant, Voice Access, and Amazon Alexa stand out, particularly in recognition accuracy and reliability, indicating their robustness. 
Meanwhile, some VAs like Voice Search (V.K.D) and Voice Search (Preeti Devi) exhibit lower performance, particularly in terms of response time and reliability.

\section{Privacy Disclosure Inconsistency}~\label{question1}
Privacy disclosures serve as the cornerstone of mobile privacy ecosystem~\cite{li2022understanding, pan2024trap, pan2024hope, tao2025privacy, wang2025big, haggag2025analysis, shanmugarasa2025privacy}. 
They are crucial in informing users about how their data is collected, processed, and shared, fostering transparency and trust. 
Numerous studies have investigated the inconsistencies between different privacy disclosure sources and actual implementation within mobile applications, including VA applications ~\cite{li2022understanding, xie2022scrutinizing, pan2023toward, pan2024trap, pan2024hope}.
As noted in this paper, while having a privacy statement is no longer a significant issue for current VAs, a new concern has emerged: some VAs show inconsistencies in their privacy statements across different platforms, or there are discrepancies between the declared privacy practices and actual usage. 
Existing studies typically focus on \textbf{bilateral inconsistencies} (e.g., only between privacy policies and privacy labels, or privacy labels and APK manifest declarations) and do not specifically examine task-executable VAs.
To fulfill the research gap, we first delve deeper into the inconsistencies and conduct a holistic empirical analysis between the following five privacy disclosure sources, namely 1) Privacy Labels (i.e. Google Play Data Safety Section), 2) Permissions identified by checker tools, 3) Permissions listed in mobile system settings, 4) Privacy policies, and 5) Declarations in manifest file. 


\begin{figure*}[t]
  \centering
   \setlength{\belowcaptionskip}{-15pt}
   \setlength{\abovecaptionskip}{-5pt}
  \includegraphics[width=\textwidth]{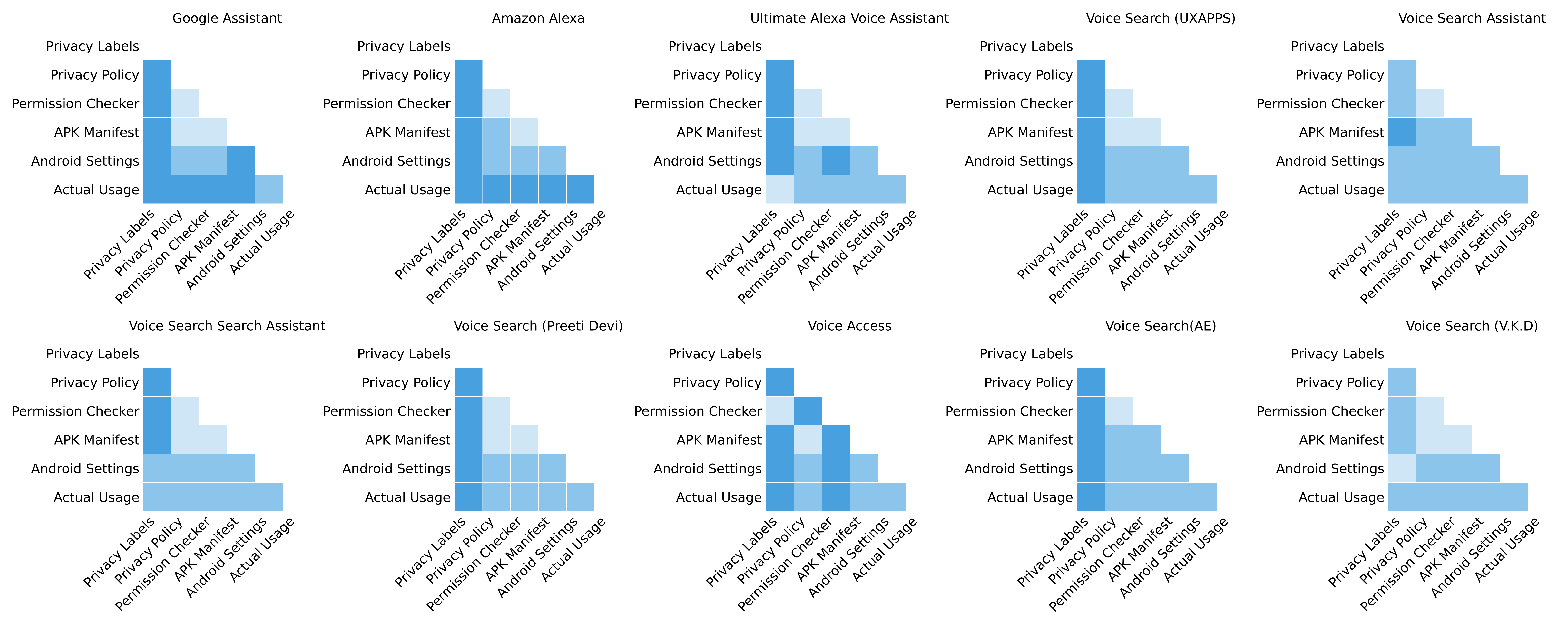} 
  \caption{The level of inconsistency for each dimension for 10 VAs. A deeper color indicates a higher degree of inconsistency in the privacy disclosures. The number denotes the count of inconsistent permissions/data types. (1) Privacy Labels. (2) Privacy Policy. (3) Permission Checker. (4) APK Manifest. (5) Android Settings. (6) Actual Usage.
        (Inconsistency level: color \colorbox{lightblue}{~} 0-2 , 
        color \colorbox{mediumblue}{~} 2-5, 
        color \colorbox{darkblue}{~} 5+).}
  \label{fig:comparison}
\end{figure*}


\textbf{(1) Privacy Labels (Data Safety Section).} 
Privacy labels provide a standardized means of informing users about an application's data handling practices~\cite{pan2023toward, li2022understanding}. For Android applications on the Google Play Store, the Data Safety Section~\cite{googleplay} serves this role by disclosing privacy-related practices in a clear and transparent manner. In our study, we manually recorded the privacy labels of the 10 examined VA applications to analyze their stated data collection and usage behaviors.

\textbf{(2) Privacy Policies}. 
Privacy policies are essential documents that outline how an application collects, uses, shares, and protects user data. 
They serve as a key mechanism for informing users about data practices and ensuring transparency and accountability~\cite{pan2024hope, pan2024trap, si2024solution, tao2025privacy}. 
For this analysis, we employed the PoliGraph tool~\cite{cui2023poligraph}, which is specifically designed to extract and structure privacy disclosures from application privacy policies. 
PoliGraph enables systematic analysis by identifying key data types and processing activities mentioned in the policies.

\textbf{(3) Permissions Identified Through Checker Tools}. 
Permission checker tools are widely used to scrutinize the permissions requested by mobile applications~\cite{felt2011android}. 
In our analysis, we utilize Permission Pilot\footnote{\href{https://github.com/d4rken-org/permission-pilot?tab=readme-ov-file}{github.com/d4rken-org/permission-pilot}}, a specialized application designed to analyze and audit the permissions declared by VA applications. 
Permission Pilot provides a detailed inspection of each app’s permission requests, offering valuable insights into the types of data and system functionalities the app can access on a user's device.

\textbf{(4) Permissions from Manifest.}  
The Android manifest is a critical XML file that provides the Android system with essential metadata about an application, including the permissions it requires to operate properly~\cite{manifest}. 
We obtain the APK from Google Plat Store and directly examine each application's manifest file (\textit{AndroidManifest.xml}) to identify all permissions explicitly declared by the app developer. 
These declared permissions represent the intended capabilities of the application and are essential for assessing its potential access to sensitive resources and functionalities.

\textbf{(5) Permissions Listed in System Settings}. 
Android applications typically display a list of granted permissions within the device's system settings. 
This list usually can be accessed under the ``Apps'' section, where users can review the specific permissions an app has been granted, such as access to location, camera, contacts, storage, etc.
Each permission corresponds to a particular capability, allowing the app to interact with various aspects of the device’s hardware or user data, such as using the microphone or reading calendar events. 
We manually record the listed permissions in the mobile system settings.

As there are different level of disclosure, some of them are per data type and the rest of them are per permissions.
The permissions are matched with permission groups \cite{mcconkey2023runtime,rahman2022permpress} as illustrated in Table \ref{tab:data-safety-permissions}, facilitating a semantic alignment between permission groups and data safety categories.
In instances where data safety types lack a direct correspondence with related Android permissions, this is denoted by ``-'' in the table. 
For example, the ```Name'' category in data safety, defined as ``How a user refers to themselves, such as their first or last name, or nickname,'' does not have an associated Android permission for users referring to themselves, resulting in no linked permission. 
Also for the permission with superset, the original permission will be picked only. For example, the \texttt{BLUETOOTH} permission allows an app to perform basic Bluetooth communication, such as connecting to paired Bluetooth devices. The \texttt{BLUETOOTH\_ADMIN permission}, on the other hand, grants more extensive control over Bluetooth functionality, including the ability to discover and pair with new devices, as well as modify Bluetooth settings. BLUETOOTH\_ADMIN is considered a superset of BLUETOOTH, meaning that an app with BLUETOOTH\_ADMIN permission implicitly has all the capabilities granted by the BLUETOOTH permission, plus additional administrative functions.
Thus we use Table \ref{tab:data-safety-permissions} to match these identified permissions with their corresponding data safety declarations.
By systematically cross-referencing permissions from these sources, we conduct fifteen unique comparisons per application, allowing us to identify and quantify the status quo of inconsistencies across the different data sources and observations.

\textbf{(6) Permission Identified Through Actual Usage.} 
We further compare examine the actual user. This data source involves a controlled testing approach to determine which permissions an Android application actually utilizes during user operations (or at least to users' impressions).
The match relation between testing commands and target data types/permissions is shown in Table~\ref{tab:data-safety-permissions}.
For example, to test the READ\_VOICEMAIL permission, the voice command ``\textit{Read voice mail}'' is used to probe the permission. 

Figure~\ref{fig:comparison} presents the comparative results of the examination. 
Due to the page limit, we do not present all the cross-checking results in the main paper but provide them in our code repository.
Our investigation reveals that privacy labels often omit permissions that are either identified through static analysis (by permission checker tools) or explicitly declared in the Android manifest files. 
For example, permissions associated with sensitive data categories, such as location or microphone access, were frequently detected in the manifest but absent from the corresponding privacy labels.
In addition, a significant misalignment was found between the permissions actively requested and those documented across various sources, including privacy statements and manifest files. 
This may suggest an over-disclosure of privacy practices, where permissions are stated but not actually used, potentially to convey a sense of compliance. 

Among the VAs studied, Google Assistant and Amazon Alexa demonstrated the greatest degree of inconsistency, whereas Voice Search (V.K.D) and Voice Search Assistant showed fewer mismatches.
This discrepancy may stem from the increased complexity and diverse functionality supported by the former.
These inconsistencies pose significant threats to user privacy by obfuscating the true extent of data access and use. The authenticity of privacy notices is a cornerstone of the notice-and-control paradigm, particularly given the autonomous nature of task-executable VAs. This concern has gained regulatory attention, as exemplified by the recent \$25 million settlement involving Apple Siri\footnote{\href{https://www.cbsnews.com/news/apple-siri-settlement-95-million-lopez-how-to-file-claim/}{www.cbsnews.com/news/apple-siri-settlement-95-million}}, underscoring the increasing scrutiny of this sector.




\section{Privacy Misdisclosure in Mega Apps}~\label{question2}
\begin{figure}[t]
    \centering
    \includegraphics[width=0.98\linewidth]{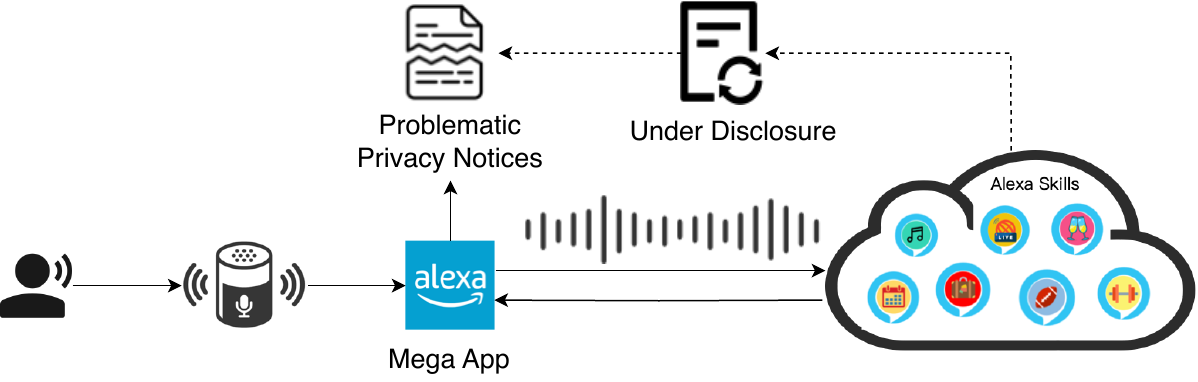}
    \caption{The under disclosure of mega apps.}
    \label{fig:MegaAppCaseStudy}
    \vspace{-13pt}
\end{figure}

During the characterization analysis mentioned in Section~\ref{section1}, we discover that many voice-based functionalities do not exist as standalone apps. 
Instead, they are packaged as mini apps that are integrated and invoked within larger, more comprehensive apps (\textit{i.e.,} mega apps). 
For example, in the case of Alexa, it provides skills to deliver voice\-based actions, including collecting health data (``\textit{Fitbit}''), making phone calls to contacts (``\textit{Calling}''), etc. 
Unfortunately, we find that these mega apps often fail to disclose all the privacy practices associated with those mini apps in their mega apps' privacy notices (e.g., Data Safety Sections).

As shown in Figure~\ref{fig:MegaAppCaseStudy}, we present a concrete example to demonstrate its implications for user privacy and the potential risks associated with undeclared permissions in Mega Apps. The diagram illustrates a scenario within the Amazon Alexa app, where voice inputs trigger the invocation of various ``Alexa skills'', allowing users to access sensitive private information, such as health data, fitness info, etc. Specifically, when a voice command \emph{``Alexa, ask Fitbit how many steps I've taken''} is issued, the fitness data is accessed and returned to the user with a response \emph{You've taken 6,000 steps today}. 
However, Alexa's privacy labels do not disclose the fitness information.

By looking into this case, we find out that there is an alarming issue where privacy declarations within skills are frequently overlooked and not adequately reflected in the corresponding mega app, leading to violations of privacy label requirements.
This oversight may not only misleads users but also leaves them uninformed about potential privacy risks, making it difficult to trace the source of data exposure in the event of a privacy breach.
This trend is not isolated to the task-executable VAs, but is similarly observed across other mega applications, such as WeChat\footnote{\url{https://www.wechat.com/}} and AliPay\footnote{\url{https://global.alipay.com/platform/site/ihome}}.
These platforms often integrate a variety of third-party functionalities (mini apps or programs), yet fail to provide sufficient visibility into the data handling practices of these components.

To address these privacy concerns, mega apps (e.g., Amazon Alexa) should adopt more transparent and comprehensive privacy notice generation practices~\cite{pan2024trap, pan2024hope, si2024solution, tao2025privacy}, covering not only the native code (i.e., code written by the app's developers) but also integrated skills and third-party libraries.
Regulatory frameworks should also enforce stricter requirements for data safety declarations, mandating that platforms clearly specify all relevant data categories. 
By improving transparency and aligning privacy declarations with actual data practices, users can be better informed about who accesses their data and for what purposes, ultimately enhancing the \textit{notice-and-control} privacy framework.

\begin{tcolorbox}[]
\textbf{Finding 1:} The privacy notices of mega apps often overlook the disclosure of privacy usage related to their mini apps (e.g., skills) or integrated functionalities. 
This lack of proper declaration can result in ineffective privacy policies, undermining user trust and potentially leading to regulatory violations.
\end{tcolorbox}

\section{The Fail of Maginot Line: Privilege Escalation Through Inter-app Interactions}~\label{question3}
In addition to the privacy disclosures caused by the mini-meta app paradigm, we discover another type of task execution pattern in
the examined VAs: inter-app interactions privilege escalation. 
Specifically, when a user sends a voice command to these VAs (the host apps), they will automatically wake up the client app to execute the action in the background, with the corresponding permissions granted to the client app, even though the host app itself lacks the necessary permission declaration.
For example, users can say ``\textit{Alexa, ask Phone Link to call Tom Brady}'', and the host app (Alexa) wakes up the client app (Phone Link) to make the phone call. 

\begin{figure}[t]
    \centering
    \includegraphics[width=0.98\linewidth]{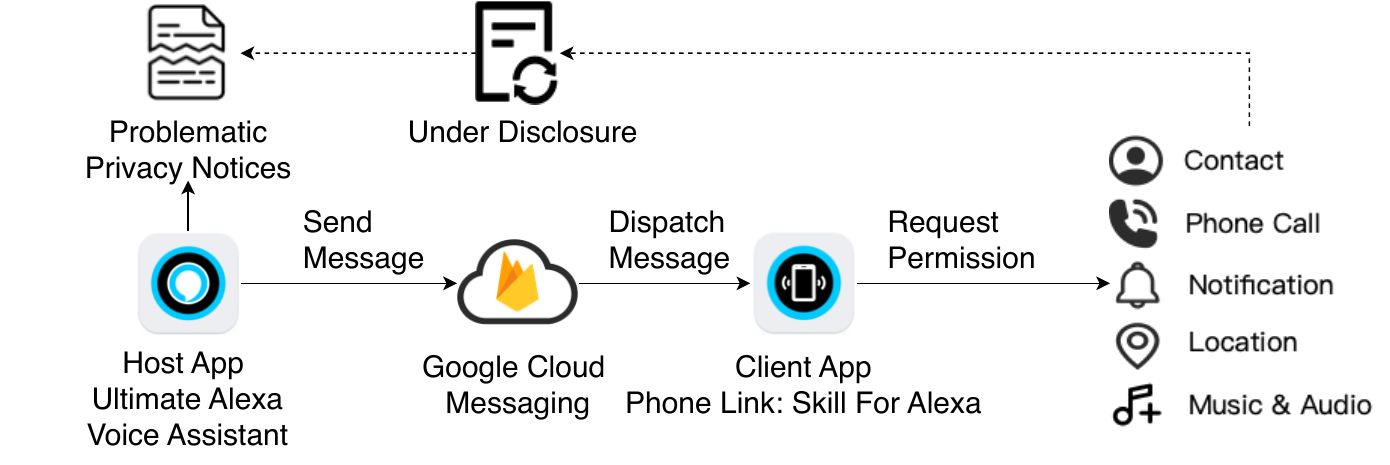}
    \caption{The mechanism of privilege escalation through inter-app interactions.}
    \label{fig:PrivilegeEscalationCaseStudy2}
    \vspace{-5pt}
\end{figure}

We then examine thesis APKs and discover that the host app does not declare any dangerous permissions, including ``android.permission.READ\_PHONE\_STATE'', ``android.permission.CALL\_PHONE'', ``android.permission. READ\_CONTACTS'', in its manifest file, yet it can still trigger sensitive behaviors by delegating them to the client app through the inter-app interactions.
This privilege escalation caused by inter-app interaction mechanism can be exploited by attackers to bypass the permission declarations by invoking dangerous behaviors from other apps, thereby leading to potential privacy risks.
We formulate the threat model as illustrated in Figure~\ref{fig:PrivilegeEscalationCaseStudy2}.

To accurately locate the epicenter of the privilege escalation and better understand the security and privacy implications, we further perform reverse engineering on the host and client APKs as shown in listing~\ref{lst:hostapp} and listing~\ref{lst:clientapp}.
From the host app's perspective, there is a \emph{BackgroundListen} (line 4 in listing~\ref{lst:hostapp}) class, within which the \emph{AvsService} is invoked (line 10 in listing~\ref{lst:hostapp}) to process audio command and then send corresponding message to cloud service (\emph{Firebase Cloud Messaging}\footnote{Firebase Cloud Messaging (FCM) is a cross-platform messaging solution that enables developers to send messages to users across different devices and platforms, including Android, iOS, and web applications : \url{https://firebase.google.com/docs/cloud-messaging}}) for further actions. This method also requests the ``android.permission.RECORD\_AUDIO'' permission (line 11 in listing~\ref{lst:hostapp}) using because it requires access to the device's microphone to capture audio. Then, on the side of the client app (listing~\ref{lst:clientapp}), a custom-designed class \emph{MessagingService} which extends \emph{FirebaseMessagingService} to handle specific intents from Firebase messages and perform corresponding actions based on those intents. Specifically, the \emph{dispatchMessage} method processes the incoming remote message and extracts the intent information from it. If the intent is ``PlaceCallInten'', it calls the \emph{makePhoneCall} method (lines 8 in listing~\ref{lst:clientapp}) to make the phone call. Inside the \emph{makePhoneCall} method, it verifies whether the app has been granted the necessary permissions \emph{READ\_PHONE\_STATE, CALL\_PHONE, READ\_CONTACTS} (lines 14-17 in listing~\ref{lst:clientapp}) and also checks if the app can draw overlays on the screen (line 19 in listing~\ref{lst:clientapp}), which is required for certain versions of Android (below API 29). Then, it retrieves contact information (lines 22-24 in listing~\ref{lst:clientapp}) and proceeds to make the phone call (lines 27 to 31 in listing~\ref{lst:clientapp}).

\begin{lstlisting}[caption={Code example demonstrating the behavior of client app. The \\ code snippet is extracted from app \emph{Phone Link: Skill For Alexa}.}, label={lst:clientapp}]
// Simplified Code Snippet of Client App
package com.customsolutions.android.phonelink;
public class MessagingService extends FirebaseMessagingService {
    ...
 public final void dispatchMessage(RemoteMessage remoteMessage) {
 //string5 is the intent info extracted from Message
 if (string5.equals("PlaceCallIntent")) {
      makePhoneCall(...);
      return;
}}

public final void makePhoneCall(...){
 ...
 if(checkSelfPermission(
 "android.permission.READ_PHONE_STATE", 
 "android.permission.CALL_PHONE", 
 "android.permission.READ_CONTACTS") 
 && 
 (Build.VERSION.SDK_INT < 29 || Settings.canDrawOverlays(...))){
     ...
     
 //Retrieve Contacts
    Cursor query = contentResolver.query(Contacts.CONTENT_URI, ...);
    Person p = new Person();
    p.phoneNumber = query.getString(...);
    ...
 //Make Phone call 
    StringBuilder sb = new StringBuilder();
    sb.append("tel:").append(p.phoneNumber);
    Intent intent = new Intent("android.intent.action.CALL",
    Uri.parse(sb.toString()));
    startActivity(intent);
};}}
\end{lstlisting}

\begin{lstlisting}[caption={Code example demonstrating the behavior of host app. The \\ code snippet is extracted from app \emph{Ultimate Alexa Voice Assistant}.}, label={lst:hostapp}]
// Simplified Code Snippet of Host App
package com.customsolutions.android.alexa;
public class BackgroundListen extends AlexaActivity{
    ...
@Override 
public void onStart() {
   ...
   //AvsService is responsible for conducting natural language data processing and sending messages to Firebase Cloud Messaging (i.e., Google Cloud Messaging).
   AvsService.sendMessage(...);
   ActivityCompat.requestPermission(this, "android.permission.RECORD_AUDIO", ...);
}}

\end{lstlisting}

Such privilege escalation, particularly when involving the misuse of dangerous permissions like \emph{android.permission.CALL\_PHONE}, poses a significant threat to users' security and privacy. 
This form of attack allows attackers to elevate their access rights, potentially enabling unauthorized actions such as making calls, accessing data, or manipulating system settings without the user’s consent. 


\begin{tcolorbox}[]
\textbf{Finding 2:}
We reveal a sophisticated privilege escalation attack model that exploits inter-app interaction mechanisms in task-executable VAs. This attack model leverages the inherent pathways between applications to escalate privileges without user consent.
\end{tcolorbox}

\section{The Abuse of Google System Applications}~\label{question4}
Based on our observations, we notice most of VAs (7/10) are either built entirely or partially built on the Google's system applications, which are pre-installed on most Android devices (e.g., Speech Recognition \& Synthesis) and use these applications for task execution.
In this section, we identify an overlooked mechanism that can be exploited by malicious VA developers: leveraging the functionalities provided by Google system applications to achieve privilege escalation.

\begin{figure}
    \centering
    \includegraphics[width=0.95\linewidth]{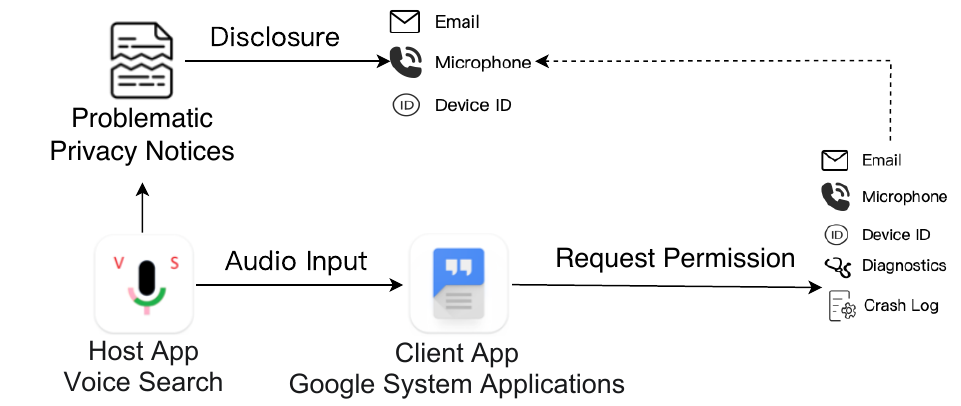}
    \caption{The mechanism of priviledge escalation through Google system application.}
    \label{fig:SystemPermissionEscalation}
\end{figure}

Specifically, as illustrated in Figure~\ref{fig:SystemPermissionEscalation}, when the user click the voice input button on ``Voice Search'', the host app wakes up the google system application ``\texttt{com.google.android.tts}''\footnote{\url{https://play.google.com/store/apps/details?id=com.google.android.tts}} to convert the audio command into text. 
Then the recognized text is returned and conduct the searching task programmed by the host app (``Voice Search'').
During this process, we suspect the host app could access and retrieve information from sensitive permissions.
Thus, we further perform reverse engineering on the host app APK and client app APK (Google Speech Recognition \& Synthesis), inspecting the code with a focus on its logic and security implications, as shown in listing~\ref{lst:voicesearch} and listing~\ref{lst:speechrecognition}. From the host application's perspective, there is a \emph{btnvoice} (line 6 in listing~\ref{lst:voicesearch}) class, within which the \emph{android.speed.action.RECOGNIZE} \emph{\_SPEECH} is invoked (line 5 in listing~\ref{lst:speechrecognition}) to make google system application proceeding the voice input. On the side of the client application (listing~\ref{lst:speechrecognition}), a custom-designed class \emph{defpackage} to handle specific intents from host application and record voice command which requiring the ``android.permission.RECORD\_AUDIO'' as it needs access to the device's microphone to capture audio.

\begin{lstlisting}[caption={Code example demonstrating the behavior of ``Voice Search''.}, label={lst:voicesearch}]
// Simplified Code Snippet of Voice Search
package com.onlinehelp3011.VoiceSearch;
public class MainActivity extends AppCompatActivity {
  ...
  // Invoke google system application for speech recognize.
  public void btnvoice(View view) {
   Intent intent = new Intent("android.speech.action.RECOGNIZE_SPEECH");
   startActivityForResult(intent, 100);
   ...
}}
\end{lstlisting}

\begin{lstlisting}[caption={Code example demonstrating the behavior of Speech \\ Recognition \& Synthesis.}, label={lst:speechrecognition}]
// Simplified Code Snippet of Speech Recognition & Synthesis (google system application)
package defpackage;
public final class fbq {
 public static ComponentName a(Context context) {
  ResolveInfo resolveInfo = packageManager().queryIntentActivities( new Intent("android.speech.action.RECOGNIZE_SPEECH");
  if (packageName.equals("com.google.android.tts") && 
      Build.VERSION.SDK_INT >= 31) {
      return new ComponentName(packageName, activityName);
}}}}
\end{lstlisting}

Additionally, detecting such data flows presents a significant challenge for program analysis, as the Google system applications involved are often closed-source and highly obfuscated. The closed-source nature and obfuscation severely impairs the ability of security analysts and automated tools to scrutinize the internal workings of these system applications, making such privacy escalation more undetectable. 

To mitigate these privacy risks, VA developers should implement more transparent data protection practices. Specifically, they need to explicitly disclose any involvement of system-provided services and clearly outline their data handling procedures. Regulatory frameworks should also impose stricter disclosure requirements to ensure users are fully informed about all entities accessing their data and the precise permissions granted.

\begin{tcolorbox}[]
\textbf{Finding 3:} 
We reveal a sophisticated privilege escalation pattern that VAs might exploit the Google system applications to avoid disclose dangerous permissions in their own privacy notices.
\end{tcolorbox}

\section{Discussions}

\begin{table}[t]
  \centering
  \caption{The tallies of 10 examined voice assistants on research questions. Black dots denote the violation exists.}
  \label{table:rqmatch}
  \resizebox{0.99\linewidth}{!}{%
  \begin{tabular}{r|c|c|c|c} 
    \toprule
    \textbf{Number} &  \makecell{Privacy Practice \\ Misdisclosure} &  \makecell{Mega-mini App \\ Interaction} &  \makecell{Inter-app \\ Interaction} &  \makecell{System Apps\\ Abuse}\\
    \midrule
    \# 1 & \filledcircle & \filledcircle & \emptycircle & \emptycircle \\ 
    \# 2 & \halfcircle & \emptycircle & \emptycircle & \emptycircle  \\
    \# 3 & \filledcircle & \filledcircle & \emptycircle & \emptycircle \\
    \# 4 & \halfcircle & \emptycircle & \filledcircle & \emptycircle \\
    \# 5 & \halfcircle & \emptycircle & \emptycircle & \filledcircle \\
    \# 6 & \emptycircle & \emptycircle & \emptycircle & \filledcircle \\
    \# 7 & \halfcircle & \emptycircle & \emptycircle & \filledcircle \\
    \# 8 & \halfcircle & \emptycircle & \emptycircle & \filledcircle \\
    \# 9 & \emptycircle & \emptycircle & \emptycircle & \filledcircle \\
    \# 10 & \halfcircle & \emptycircle & \emptycircle & \filledcircle \\
  \bottomrule
\end{tabular}
}%
\end{table}

Based on the analysis of the 10 selected VAs, Table~\ref{table:rqmatch} shows the tallies of observed privacy risks to the four aspects we discussed. 
Most of the VAs exhibited at least one form of privacy violation, indicating prevalent privacy concerns in current VA ecosystem.
Specifically, privacy practice misdisclosures emerged as the most common issue, affecting half of the examined VAs. 
Mega-mini app interaction issues also appeared prominently, impacting four assistants, while inter-app interactions and system app abuses were less common but still significant, observed in one and two assistants, respectively. 
These results highlight that privacy risks are widely distributed across different categories.
Based on our observations and study results, we summarize some findings for various roles or stakeholders in the ecosystem.

\textbf{App Developers.} 
Developers of VAs must exercise heightened diligence regarding privacy disclosure, particularly within their privacy labels. 
Our findings underscore a persistent divergence between declared data practices and actual permission usage, signaling a need for greater transparency. 
Since the privacy labels and policies often constitute users' first point of contact with an application, its contents must be comprehensive, accurate, and reflective of the true data access behaviors of the application. 
Additionally, developers should be careful on employing inter-application interactions or leveraging Google system applications as a means of bypassing explicit permission declarations. 
Ensuring such integrity is vital to preserving user trust and maintaining compliance with platform policies and privacy regulations and laws.

\textbf{Privacy regulators.} 
Regulatory bodies should be aware of the distinctive privacy risks introduced by task-executable VAs. 
The current oversight mechanisms insufficiently capture the complex and often opaque data flows facilitated by these apps, especially those involving integrated third-party functionalities and system-level privilege escalation. 
To remediate these gaps, regulators should develop targeted frameworks that mandate exhaustive and transparent permission disclosures, encompassing all forms of data access, including those mediated by system applications or inter-app communication channels. 
Furthermore, regulations should explicitly address privacy governance within mega applications, requiring clear demarcation and disclosure of data practices related to mini apps or mini programs.
Such measures are essential for safeguarding user autonomy and ensuring accountability within the voice assistant ecosystem.

\section{Implications to Autonomous AI Agents}
There is an increasing number of autonomous AI agents emerging every day, claiming capabilities that extend far beyond traditional \textit{``voice assistants.''} 
These advanced AI agents, mostly LLM-driven sophisticated systems, perform various complex tasks on user devices, often involving sensitive personal data and critical system functions. 
The privacy risks observed in our analysis of VAs can similarly apply to these autonomous AI agents, potentially exacerbating existing vulnerabilities and introducing new privacy threats.
Notably, the cross-app operational model of autonomous AI agents increases the likelihood of privilege escalation and circumvention of traditional permission-based privacy frameworks. Through inter-app interactions or implicit invocation of system-level services, these agents may gain unauthorized access to protected resources, mirroring the threat models uncovered in our study

Moreover, the inference capabilities of LLMs further heighten risks through permission-based vulnerabilities and potential side-channel attacks.
For example, an LLM can easily infer the demographic information (e.g., ethnicity, address) based on the restaurants you ask the agent to book recursively.
Therefore, the identified privacy issues should prompt stakeholders of autonomous AI agents to adopt rigorous privacy standards, implement transparent disclosure practices, and ensure more robust mechanisms to mitigate the potential risks highlighted by our findings.

\section{Related Work}

\subsection{Privacy Compliance of VAs}
People have broadly studied the privacy issues in VAs.
Xie et al. \cite{xie2022scrutinizing} developed Skipper, a tool that automatically analyzes privacy policies of Amazon Alexa skills using NLP and machine learning techniques to detect non-compliance issues. Yan et al. \cite{yan2024quality} proposed QuPer, which assesses privacy policies based on four metrics: timeliness, availability, completeness, and readability. Liu et al. \cite{li2024model} introduced Elevate, a model-enhanced LLM-driven VUI testing framework for VPA apps. It combines LLMs with behavioral models for improved testing coverage. Liao et al. \cite{liao2024understanding,liao2024ensuring} conducted comprehensive studies on privacy compliance in VPA apps, including GDPR compliance in European marketplaces and large-scale analysis of privacy policies in US marketplaces. 
They developed SkillScanner, a static code analysis tool to detect privacy and policy violations in skill development, and SkillPoV, a tool for generating voice-based privacy notices.
Nevertheless, the risk and challenges associated with greater autonomy introduced by the task-execution VAs are rarely noticed.

\subsection{IoT Security and Privacy Risks}

Alrawi et al. \cite{alrawi2019sok} conducted a comprehensive survey of IoT security, proposing a four-pronged classification model for analyzing IoT security. Sivaraman et al. \cite{sivaraman2015network} explored network-level security for smart home IoT devices, highlighting the importance of securing the communication between devices. Zhang et al. \cite{zhang2018homonit} investigated privacy issues in smart homes, focusing on information leakage through encrypted IoT traffic. In the context of voice-controlled IoT, Ding et al. \cite{ding2024command} discovered new vulnerabilities in the Amazon Alexa platform, demonstrating how malicious third-party developers could hijack built-in voice commands to invoke malicious IoT skills. 

\subsection{Android Program Analysis} Our fellow researchers
have proposed various approaches to perform program analysis on Android apps~\cite{sun2021taming, samhi2022jucify,arzt2014flowdroid, sun2022demystifying, sun2022mining, zhou2024bridging, zhao2022code, sun2023taming, liu2022first, chen2024your, sun2023lazycow, liu2025comparative, hu2023detecting}. For example, Arzt et al. \cite{arzt2014flowdroid} developed FlowDroid, a precise context, flow, field, and object-sensitive taint analysis tool for Android apps. These techniques have been applied to detect various security issues, including over-privilege \cite{felt2011android}, unauthorized data access \cite{zhou2012hey}, insecure data transmission \cite{fahl2012eve}, and component hijacking \cite{chin2011analyzing}. 
While existing studies provide significant insights into privacy and security risks, they often overlook the unique challenges posed by task-executable voice assistants. 
Our research addresses this gap by offering a timely and comprehensive investigation into these challenges.

\section{Conclusion}

Task-executable voice assistants (VAs) have become more popular, enhancing user convenience and expanding device functionality.
Android task-executable VAs are applications that are capable of understanding complex tasks and performing corresponding operations.
In this paper, we presented a user-centric comprehensive empirical study on privacy risks in Android task-executable VA applications.
We collected ten mainstream VAs as our research target and analyze their operational characteristics. 
We then cross-checked their privacy declarations across six sources, including privacy labels, policies, and manifest files, and our findings revealed widespread inconsistencies.
Moreover, we uncovered three significant privacy threat models: (1) privacy misdisclosure in mega apps, where integrated mini apps such as Alexa skills are inadequately represented; (2) privilege escalation via inter-application interactions, which exploit Android's communication mechanisms to bypass user consent; and (3) abuse of Google system applications, enabling apps to evade the declaration of dangerous permissions.
This study contributes actionable recommendations for practitioners and underscores the broader relevance of these privacy risks to emerging autonomous AI agents.

\noindent \textbf{Data Availability.}
The link to our repository is: \url{https://github.com/ShidongPAN/Task_VA}

\bibliographystyle{IEEEtran}
\bibliography{bibliography}

\begin{thebibliography}{10}
\providecommand{\url}[1]{#1}
\csname url@samestyle\endcsname
\providecommand{\newblock}{\relax}
\providecommand{\bibinfo}[2]{#2}
\providecommand{\BIBentrySTDinterwordspacing}{\spaceskip=0pt\relax}
\providecommand{\BIBentryALTinterwordstretchfactor}{4}
\providecommand{\BIBentryALTinterwordspacing}{\spaceskip=\fontdimen2\font plus
\BIBentryALTinterwordstretchfactor\fontdimen3\font minus \fontdimen4\font\relax}
\providecommand{\BIBforeignlanguage}[2]{{%
\expandafter\ifx\csname l@#1\endcsname\relax
\typeout{** WARNING: IEEEtran.bst: No hyphenation pattern has been}%
\typeout{** loaded for the language `#1'. Using the pattern for}%
\typeout{** the default language instead.}%
\else
\language=\csname l@#1\endcsname
\fi
#2}}
\providecommand{\BIBdecl}{\relax}
\BIBdecl

\bibitem{murad2019revolution}
C.~Murad, C.~Munteanu, B.~R. Cowan, and L.~Clark, ``Revolution or evolution? speech interaction and hci design guidelines,'' \emph{IEEE Pervasive Computing}, vol.~18, no.~2, pp. 33--45, 2019.

\bibitem{clark2019state}
L.~Clark, P.~Doyle, D.~Garaialde, E.~Gilmartin, S.~Schl{\"o}gl, J.~Edlund, M.~Aylett, J.~Cabral, C.~Munteanu, J.~Edwards \emph{et~al.}, ``The state of speech in hci: Trends, themes and challenges,'' \emph{Interacting with computers}, vol.~31, no.~4, pp. 349--371, 2019.

\bibitem{bolton2021security}
T.~Bolton, T.~Dargahi, S.~Belguith, M.~S. Al-Rakhami, and A.~H. Sodhro, ``On the security and privacy challenges of virtual assistants,'' \emph{Sensors}, vol.~21, no.~7, p. 2312, 2021.

\bibitem{acosta2022survey}
L.~H. Acosta and D.~Reinhardt, ``A survey on privacy issues and solutions for voice-controlled digital assistants,'' \emph{Pervasive and Mobile Computing}, vol.~80, p. 101523, 2022.

\bibitem{liao2020measuring}
S.~Liao, C.~Wilson, L.~Cheng, H.~Hu, and H.~Deng, ``Measuring the effectiveness of privacy policies for voice assistant applications,'' in \emph{Proceedings of the 36th Annual Computer Security Applications Conference}, 2020, pp. 856--869.

\bibitem{wu2025assistants}
L.~Wu, C.~Wang, T.~Liu, Y.~Zhao, and H.~Wang, ``From assistants to adversaries: Exploring the security risks of mobile llm agents,'' \emph{arXiv preprint arXiv:2505.12981}, 2025.

\bibitem{lin2025mind}
Z.~Lin, J.~Li, S.~Pan, Y.~Shi, Y.~Yao, and D.~Xu, ``Mind the third eye! benchmarking privacy awareness in mllm-powered smartphone agents,'' \emph{arXiv preprint arXiv:2508.19493}, 2025.

\bibitem{bbcamazon2023}
\BIBentryALTinterwordspacing
G.~Wright. Amazon to pay \$25m over child privacy violations. [Online]. Available: \url{https://www.bbc.com/news/technology-65772154}
\BIBentrySTDinterwordspacing

\bibitem{natatsuka2019poster}
A.~Natatsuka, R.~Iijima, T.~Watanabe, M.~Akiyama, T.~Sakai, and T.~Mori, ``Poster: A first look at the privacy risks of voice assistant apps,'' in \emph{Proceedings of the 2019 ACM SIGSAC Conference on Computer and Communications Security}, 2019, pp. 2633--2635.

\bibitem{xie2022scrutinizing}
F.~Xie, Y.~Zhang, C.~Yan, S.~Li, L.~Bu, K.~Chen, Z.~Huang, and G.~Bai, ``Scrutinizing privacy policy compliance of virtual personal assistant apps,'' in \emph{Proceedings of the 37th IEEE/ACM international conference on automated software engineering}, 2022, pp. 1--13.

\bibitem{terzopoulos2020voice}
G.~Terzopoulos and M.~Satratzemi, ``Voice assistants and smart speakers in everyday life and in education,'' \emph{Informatics in Education}, vol.~19, no.~3, pp. 473--490, 2020.

\bibitem{hoy2018alexa}
M.~B. Hoy, ``Alexa, siri, cortana, and more: an introduction to voice assistants,'' \emph{Medical reference services quarterly}, vol.~37, no.~1, pp. 81--88, 2018.

\bibitem{kepuska2018next}
V.~Kepuska and G.~Bohouta, ``Next-generation of virtual personal assistants (microsoft cortana, apple siri, amazon alexa and google home),'' in \emph{2018 IEEE 8th annual computing and communication workshop and conference (CCWC)}.\hskip 1em plus 0.5em minus 0.4em\relax IEEE, 2018, pp. 99--103.

\bibitem{cheng2022personal}
P.~Cheng and U.~Roedig, ``Personal voice assistant security and privacy—a survey,'' \emph{Proceedings of the IEEE}, vol. 110, no.~4, pp. 476--507, 2022.

\bibitem{pan2023toward}
S.~Pan, T.~Hoang, D.~Zhang, Z.~Xing, X.~Xu, Q.~Lu, and M.~Staples, ``Toward the cure of privacy policy reading phobia: Automated generation of privacy nutrition labels from privacy policies,'' \emph{arXiv preprint arXiv:2306.10923}, 2023.

\bibitem{pan2024hope}
\BIBentryALTinterwordspacing
S.~Pan, Z.~Tao, T.~Hoang, D.~Zhang, T.~Li, Z.~Xing, X.~Xu, M.~Staples, T.~Rakotoarivelo, and D.~Lo, ``A {NEW} {HOPE}: Contextual privacy policies for mobile applications and an approach toward automated generation,'' in \emph{33rd USENIX Security Symposium (USENIX Security 24)}.\hskip 1em plus 0.5em minus 0.4em\relax Philadelphia, PA: USENIX Association, Aug. 2024, pp. 5699--5716. [Online]. Available: \url{https://www.usenix.org/conference/usenixsecurity24/presentation/pan-shidong-hope}
\BIBentrySTDinterwordspacing

\bibitem{pan2024trap}
\BIBentryALTinterwordspacing
S.~Pan, D.~Zhang, M.~Staples, Z.~Xing, J.~Chen, X.~Xu, and T.~Hoang, ``Is it a trap? a large-scale empirical study and comprehensive assessment of online automated privacy policy generators for mobile apps,'' in \emph{33rd USENIX Security Symposium (USENIX Security 24)}.\hskip 1em plus 0.5em minus 0.4em\relax Philadelphia, PA: USENIX Association, Aug. 2024, pp. 5681--5698. [Online]. Available: \url{https://www.usenix.org/conference/usenixsecurity24/presentation/pan-shidong-trap}
\BIBentrySTDinterwordspacing

\bibitem{lau2018alexa}
J.~Lau, B.~Zimmerman, and F.~Schaub, ``Alexa, are you listening? privacy perceptions, concerns and privacy-seeking behaviors with smart speakers,'' \emph{Proceedings of the ACM on human-computer interaction}, vol.~2, no. CSCW, pp. 1--31, 2018.

\bibitem{huang2020amazon}
Y.~Huang, B.~Obada-Obieh, and K.~Beznosov, ``Amazon vs. my brother: How users of shared smart speakers perceive and cope with privacy risks,'' in \emph{Proceedings of the 2020 CHI conference on human factors in computing systems}, 2020, pp. 1--13.

\bibitem{alrawi2019sok}
O.~Alrawi, C.~Lever, M.~Antonakakis, and F.~Monrose, ``Sok: Security evaluation of home-based iot deployments,'' in \emph{2019 IEEE symposium on security and privacy (sp)}.\hskip 1em plus 0.5em minus 0.4em\relax IEEE, 2019, pp. 1362--1380.

\bibitem{liao2024understanding}
S.~Liao, M.~Aldeen, J.~Yan, L.~Cheng, X.~Luo, H.~Cai, and H.~Hu, ``Understanding gdpr non-compliance in privacy policies of alexa skills in european marketplaces,'' in \emph{Proceedings of the ACM on Web Conference 2024}, 2024, pp. 1081--1091.

\bibitem{li2022understanding}
T.~Li, K.~Reiman, Y.~Agarwal, L.~F. Cranor, and J.~I. Hong, ``Understanding challenges for developers to create accurate privacy nutrition labels,'' in \emph{Proceedings of the 2022 CHI Conference on Human Factors in Computing Systems}.

\bibitem{tao2025privacy}
Z.~Tao, S.~Pan, Z.~Xing, X.~Sun, O.~Haggag, J.~Grundy, J.~Li, and L.~Zhu, ``Privacy bills of materials (pribom): A transparent privacy information inventory for collaborative privacy notice generation in mobile app development,'' in \emph{The 25th Privacy Enhancing Technologies Symposium}.\hskip 1em plus 0.5em minus 0.4em\relax Privacy Enhancing Technologies Board, 2025, pp. 392--409.

\bibitem{wang2025big}
L.~Wang, D.~Wang, S.~Pan, Z.~Jiang, H.~Wang, and Y.~Wang, ``A big step forward? a user-centric examination of ios app privacy report and enhancements,'' in \emph{2025 IEEE Symposium on Security and Privacy (SP)}.\hskip 1em plus 0.5em minus 0.4em\relax IEEE, 2025, pp. 4210--4228.

\bibitem{haggag2025analysis}
O.~Haggag, A.~Pedace, S.~Pan, and J.~Grundy, ``An analysis of privacy regulations and user concerns of finance mobile applications,'' \emph{Information and Software Technology}, p. 107756, 2025.

\bibitem{shanmugarasa2025privacy}
Y.~Shanmugarasa, S.~Pan, M.~Ding, D.~Zhao, and T.~Rakotoarivelo, ``Privacy meets explainability: Managing confidential data and transparency policies in llm-empowered science,'' in \emph{Proceedings of the Extended Abstracts of the CHI Conference on Human Factors in Computing Systems}, 2025, pp. 1--8.

\bibitem{googleplay}
\BIBentryALTinterwordspacing
G.~Play. Provide information for google play's data safety section. [Online]. Available: \url{https://support.google.com/googleplay/android-developer/answer/10787469?hl=en}
\BIBentrySTDinterwordspacing

\bibitem{si2024solution}
M.~Si, S.~Pan, D.~Liao, X.~Sun, Z.~Tao, W.~Shi, and Z.~Xing, ``A solution toward transparent and practical ai regulation: Privacy nutrition labels for open-source generative ai-based applications,'' \emph{arXiv preprint arXiv:2407.15407}, 2024.

\bibitem{cui2023poligraph}
H.~Cui, R.~Trimananda, A.~Markopoulou, and S.~Jordan, ``$\{$PoliGraph$\}$: Automated privacy policy analysis using knowledge graphs,'' in \emph{32nd USENIX Security Symposium (USENIX Security 23)}, 2023, pp. 1037--1054.

\bibitem{felt2011android}
A.~P. Felt, E.~Chin, S.~Hanna, D.~Song, and D.~Wagner, ``Android permissions demystified,'' in \emph{Proceedings of the 18th ACM conference on Computer and communications security}, 2011, pp. 627--638.

\bibitem{manifest}
\BIBentryALTinterwordspacing
Android. App manifest overview. [Online]. Available: \url{https://developer.android.com/guide/topics/manifest/manifest-intro}
\BIBentrySTDinterwordspacing

\bibitem{mcconkey2023runtime}
R.~McConkey and O.~Olukoya, ``Runtime and design time completeness checking of dangerous android app permissions against gdpr,'' \emph{IEEE Access}, 2023.

\bibitem{rahman2022permpress}
M.~S. Rahman, P.~Naghavi, B.~Kojusner, S.~Afroz, B.~Williams, S.~Rampazzi, and V.~Bindschaedler, ``Permpress: Machine learning-based pipeline to evaluate permissions in app privacy policies,'' \emph{IEEE Access}, vol.~10, pp. 89\,248--89\,269, 2022.

\bibitem{yan2024quality}
C.~Yan, F.~Xie, M.~H. Meng, Y.~Zhang, and G.~Bai, ``On the quality of privacy policy documents of virtual personal assistant applications,'' \emph{Proceedings on Privacy Enhancing Technologies}, 2024.

\bibitem{li2024model}
S.~Li, L.~Bu, G.~Bai, F.~Xie, K.~Chen, and C.~Yue, ``Model-enhanced llm-driven vui testing of vpa apps,'' \emph{arXiv preprint arXiv:2407.02791}, 2024.

\bibitem{liao2024ensuring}
S.~Liao, ``Ensuring the privacy compliance of voice personal assistant applications,'' 2024.

\bibitem{sivaraman2015network}
V.~Sivaraman, H.~H. Gharakheili, A.~Vishwanath, R.~Boreli, and O.~Mehani, ``Network-level security and privacy control for smart-home iot devices,'' in \emph{2015 IEEE 11th International conference on wireless and mobile computing, networking and communications (WiMob)}.\hskip 1em plus 0.5em minus 0.4em\relax IEEE, 2015, pp. 163--167.

\bibitem{zhang2018homonit}
W.~Zhang, Y.~Meng, Y.~Liu, X.~Zhang, Y.~Zhang, and H.~Zhu, ``Homonit: Monitoring smart home apps from encrypted traffic,'' in \emph{Proceedings of the 2018 ACM SIGSAC Conference on Computer and Communications Security}, 2018, pp. 1074--1088.

\bibitem{ding2024command}
W.~Ding, S.~Liao, L.~Cheng, X.~Mi, Z.~Zhao, and H.~Hu, ``Command hijacking on voice-controlled iot in amazon alexa platform,'' in \emph{Proceedings of the 19th ACM Asia Conference on Computer and Communications Security}, 2024, pp. 654--666.

\bibitem{sun2021taming}
X.~Sun, L.~Li, T.~F. Bissyand{\'e}, J.~Klein, D.~Octeau, and J.~Grundy, ``Taming reflection: An essential step toward whole-program analysis of android apps,'' \emph{ACM Transactions on Software Engineering and Methodology (TOSEM)}, vol.~30, no.~3, pp. 1--36, 2021.

\bibitem{samhi2022jucify}
J.~Samhi, J.~Gao, N.~Daoudi, P.~Graux, H.~Hoyez, X.~Sun, K.~Allix, T.~F. Bissyand{\'e}, and J.~Klein, ``Jucify: A step towards android code unification for enhanced static analysis,'' in \emph{Proceedings of the 44th International Conference on Software Engineering}, 2022, pp. 1232--1244.

\bibitem{arzt2014flowdroid}
S.~Arzt, S.~Rasthofer, C.~Fritz, E.~Bodden, A.~Bartel, J.~Klein, Y.~Le~Traon, D.~Octeau, and P.~McDaniel, ``Flowdroid: Precise context, flow, field, object-sensitive and lifecycle-aware taint analysis for android apps,'' \emph{ACM sigplan notices}, vol.~49, no.~6, pp. 259--269, 2014.

\bibitem{sun2022demystifying}
X.~Sun, X.~Chen, L.~Li, H.~Cai, J.~Grundy, J.~Samhi, T.~F. Bissyand{\'e}, and J.~Klein, ``Demystifying hidden sensitive operations in android apps,'' \emph{ACM Transactions on Software Engineering and Methodology}, 2022.

\bibitem{sun2022mining}
X.~Sun, X.~Chen, Y.~Zhao, P.~Liu, J.~Grundy, and L.~Li, ``Mining android api usage to generate unit test cases for pinpointing compatibility issues,'' in \emph{Proceedings of the 37th IEEE/ACM International Conference on Automated Software Engineering}, 2022, pp. 1--13.

\bibitem{zhou2024bridging}
T.~Zhou, Y.~Zhao, X.~Hou, X.~Sun, K.~Chen, and H.~Wang, ``Bridging design and development with automated declarative ui code generation,'' \emph{arXiv preprint arXiv:2409.11667}, 2024.

\bibitem{zhao2022code}
Y.~Zhao, L.~Li, X.~Sun, P.~Liu, and J.~Grundy, ``Code implementation recommendation for android gui components,'' in \emph{Proceedings of the ACM/IEEE 44th International Conference on Software Engineering: Companion Proceedings}, 2022, pp. 31--35.

\bibitem{sun2023taming}
X.~Sun, X.~Chen, Y.~Liu, J.~Grundy, and Li, ``Taming android fragmentation through lightweight crowdsourced testing,'' \emph{IEEE Transactions on Software Engineering}, 2023.

\bibitem{liu2022first}
P.~Liu, X.~Sun, Y.~Zhao, Y.~Liu, J.~Grundy, and L.~Li, ``A first look at ci/cd adoptions in open-source android apps,'' in \emph{Proceedings of the 37th IEEE/ACM International Conference on Automated Software Engineering}, 2022, pp. 1--6.

\bibitem{chen2024your}
H.~Chen, D.~Chen, Y.~Liu, X.~Sun, and L.~Li, ``Are your android app analyzers still relevant?'' in \emph{Proceedings of the IEEE/ACM 11th International Conference on Mobile Software Engineering and Systems}, 2024, pp. 69--73.

\bibitem{sun2023lazycow}
X.~Sun, X.~Chen, Y.~Liu, J.~Grundy, and L.~Li, ``Lazycow: A lightweight crowdsourced testing tool for taming android fragmentation,'' in \emph{Proceedings of the 31st ACM Joint European Software Engineering Conference and Symposium on the Foundations of Software Engineering}, 2023, pp. 2127--2131.

\bibitem{liu2025comparative}
Y.~Liu, X.~Chen, Y.~Liu, P.~Kong, T.~F. Bissyand{\'e}, J.~Klein, X.~Sun, L.~Li, C.~Chen, and J.~Grundy, ``A comparative study between android phone and tv apps,'' \emph{Automated Software Engineering}, vol.~32, no.~2, p.~46, 2025.

\bibitem{hu2023detecting}
H.~Hu, Y.~Liu, Y.~Zhao, Y.~Liu, X.~Sun, C.~Tantithamthavorn, and L.~Li, ``Detecting temporal inconsistency in biased datasets for android malware detection,'' in \emph{2023 38th IEEE/ACM International Conference on Automated Software Engineering Workshops (ASEW)}.\hskip 1em plus 0.5em minus 0.4em\relax IEEE, 2023, pp. 17--23.

\bibitem{zhou2012hey}
Y.~Zhou, Z.~Wang, W.~Zhou, and X.~Jiang, ``Hey, you, get off of my market: detecting malicious apps in official and alternative android markets.'' in \emph{NDSS}, vol.~25, no.~4, 2012, pp. 50--52.

\bibitem{fahl2012eve}
S.~Fahl, M.~Harbach, T.~Muders, L.~Baumg{\"a}rtner, B.~Freisleben, and M.~Smith, ``Why eve and mallory love android: An analysis of android ssl (in) security,'' in \emph{Proceedings of the 2012 ACM conference on Computer and communications security}, 2012, pp. 50--61.

\bibitem{chin2011analyzing}
E.~Chin, A.~P. Felt, K.~Greenwood, and D.~Wagner, ``Analyzing inter-application communication in android,'' in \emph{Proceedings of the 9th international conference on Mobile systems, applications, and services}, 2011, pp. 239--252.

\end{thebibliography}

\end{document}